# As long as you talk about me: The importance of family firm brands and the contingent role of family-firm identity

Rovelli, P., Benedetti, C., Fronzetti Colladon, A., & De Massis, A.

# As long as you talk about me:

## The importance of family firm brands and the contingent role of family-firm identity

*"There is only one thing in the world worse than being talked about,*

*and that is not being talked about"*

(The Picture of Dorian Gray, 1890)

**ABSTRACT**

This study explores the role of external audiences in determining the importance of family firm brands and the relationship with firm performance. Drawing on text mining and social network analysis techniques, and considering the brand prevalence, diversity, and connectivity dimensions, we use the semantic brand score to measure the importance the media give to family firm brands. The analysis of a sample of 52,555 news articles published in 2017 about 63 Italian entrepreneurial families reveals that brand importance is positively associated with family firm revenues, and this relationship is stronger when there is identity match between the family and the firm. This study advances current literature by offering a rich and multifaceted perspective on how external audiences perceptions of the brand shape family firm performance.

# 1. Introduction

Studying the branding strategies of family firms is critical to understand how they communicate at the intersection of two idiosyncratic systems, the family and the business. Research studying family firm brands has mainly adopted an internal perspective, exploring how family firms strategically manage their brands (Botero, Thomas, Graves, & Fediuk, 2013; Micelotta & Raynard, 2011), and how strategic decisions influence brand relevance (Ardito, Messeni Petruzzelli, Pascucci, & Peruffo, 2019; Magistretti, Dell'Era, Frattini, & Petruzzelli, 2020; Mazzelli, De Massis, Messeni Petruzzelli, Del Giudice, & Khan, 2020). These studies show that branding strategies that communicate the family nature of the firm positively relate to firm performance (Zellweger, Kellermanns, Eddleston, & Memili, 2012b), and that viewing the family "as a corporate brand" leads to higher sales growth. However, brand management concerns not only the party communicating or transmitting the informational cues pertaining to the brand but also the party receiving and processing such information (Brown, Dacin, Pratt, & Whetten, 2006; Shannon, 1948). Therefore, adopting an external perspective of family firm brands and the importance that external stakeholders attribute to them is crucial to advance our understanding of how these stakeholders can affect consumer choices that lead to the firm's competitive advantage.

In this respect, a relatively less explored research stream acknowledges the role of external audiences in determining the importance of family firm brands (Binz, Hair, Pieper, & Baldauf, 2013; Orth & Green, 2009; Sageder, Duller, & Mitter, 2015), and how this may ultimately impact firm performance. Moreover, research adopting an external perspective of family firm brands predominantly focuses on consumers, disregarding the heterogeneity among different external stakeholders and their informational cues (Beck, 2016). Extending this narrow body of knowledge requires digging deeper into the role of the media as authoritative sources of information that have the power to influence stakeholder perceptions and opinions (Deephouse, 2000; Hoffman & Ocasio, 2001). To this end, we study media narratives, taking into account how they promote family firm brands, improving their awareness, through embeddedness in a rich and distinctive discourse, and their connection with different topics. We rely on a recent conceptualization of *brand importance* (Fronzetti Colladon, 2018) using the semantic brand score (SBS) brand intelligence (BI) tool (Fronzetti Colladon & Grippa, 2020) to evaluate the degree of importance that external stakeholders attribute to a brand. Specifically, we use the SBS composite indicator comprising the three brand importance dimensions of prevalence, diversity, and connectivity. Although partly grounded in well-known brand equity models (e.g., Keller, 1993), this approach extends the range of possible analyses and provides additional information through exploring big news data.

Studies have analyzed brand prevalence in terms of media coverage as a factor affecting individual choices and preferences (e.g., Liu & Lopez, 2016), yet there is evidence that just looking at how frequently a brand name is mentioned is not sufficient to fully capture the magnitude of its potential impact (Fronzetti Colladon, 2020). While media visibility is a proxy of brand prevalence that can increase brand awareness, i.e., recognition and recall (Keller, 1993), we should also consider the brand image that news articles convey through specific associations with the brand. In particular, to more fully evaluate brand importance, the concept of prevalence has to be complemented with the brand diversity and connectivity constructs. Diversity looks at the richness and uniqueness of brand association, with evidence of a positive impact on brand strength (Grohs, Raies, Koll, & Mühlbacher, 2016). Indeed, a brand name might appear frequently, but in a very narrow discourse, and thus be of limited importance. Last, connectivity points to the brand's "brokerage power", i.e., its ability to potentially connect different words and/or discourse topics. These three dimensions together constitute the brand importance construct (Fronzetti Colladon, 2018).

Accordingly, we aim to offer a new perspective of the management of family firm brands by exploring whether brand importance is related to performance in terms of revenues. Moreover, while adopting an external perspective is important to advance marketing and branding research in family business, we cannot ignore the nature of these firms. Specifically, we contend that the relationship between brand importance and revenues might be contingent on a distinctive characteristic of family firms – namely, the family's identification with the firm (e.g., Chrisman, Chua, & Sharma, 2005; De Massis, Kotlar, Mazzola, Minola, & Sciascia, 2018; Zellweger, Eddleston, & Kellermanns, 2010). As such, we investigate whether the relationship between brand importance and firm revenues changes depending on the family-firm identity. In summary, we aim to answer the following research questions: *How does the brand importance attributed by external audiences relate to family firms' revenues? What role does the family-firm identity play in this respect?*

To this aim, we combine social network and semantic analysis methods to analyze online news data and reveal insightful information and trends. In particular, we use our approach on a sample of 52,555 news articles published in 2017 related to 63 Italian family firms. Taking inspiration from previous studies (e.g., Micelotta & Raynard, 2011), we selected our sample from the Forbes 2018 ranking of the Top 100 Italian entrepreneurial families and their businesses.

Our findings reveal that family firms obtain a competitive advantage in terms of revenues when brand importance is conveyed by external audiences. This effect is stronger when there is an overlap between the family and the firm name, namely when the family identifies with the firm. Consumers' emotional bonds, affect, and orientation toward the brand resulting from the family firm highlighting the familial component and promoting the family background enhance the positive effect of brand importance derived from external audiences.

This study contributes to the family business literature by adopting an external perspective of family firm brands, leveraging the concept of brand importance to delineate how external stakeholders perceive such firms. Moving forward from the dominant consumers' view, we incorporate a media perspective of family firm brands, shedding new light on external stakeholder heterogeneity. Furthermore, we go beyond the use of traditional approaches – such as surveys, case studies, interviews, and focus groups – to study family firms by drawing on big data and semantic network analysis.

## 2. Theoretical background and hypotheses

Scholars have recently started paying more attention to family firms' marketing and branding strategies, as testified by the proliferation of literature reviews on this topic (Andreini, Bettinelli, Pedeliento, & Apa, 2020; Astrachan, Botero, Astrachan, & Prügl, 2018; Beck, 2016; Bravo, Cambra,

Centeno, & Melero, 2017; Sageder, Mitter, & Feldbauer-Durstmüller, 2018). In particular, family business scholars have shown interest in strategic decisions linked to branding the firm as a *family firm* (Kashmiri & Mahajan, 2010; Micelotta & Raynard, 2011), revealing that the brand can become an important source of differentiation among organizations (Astrachan et al., 2018). Most family business studies on branding adopt an internal organizational perspective, exploring how family firms strategically manage their brands at the intersection of the family and business systems (Botero et al., 2013; Micelotta & Raynard, 2011), and how strategic decisions influence brand relevance (Ardito et al., 2019; Magistretti et al., 2020; Mazzelli et al., 2020). However, only a few studies look at stakeholder perceptions of family firms, relying mainly on survey data and experimental research (e.g., Beck & Kenning, 2015; Carrigan & Buckley, 2008; Lude & Prügl, 2018; Schellong, Kraiczy, Malär, & Hack, 2019).

In our view, these approaches do not enable fully understanding family firm brands and strategies, and rather neglect the opportunities afforded by the availability of rich online text data. Indeed, the increasing accessibility to big textual data from multiple media and social media sources offers scholars the opportunity to study the value and importance of brands considering the opinions and expressions of external stakeholders (Fronzetti Colladon, 2018; Kim & Ko, 2012; Tsimonis & Dimitriadis, 2014). Thanks to big data analytics, new patterns and trends have emerged in the study of brands. This enriched knowledge can drive strategic choices and support novel family firm branding strategies and actions (Kunz et al., 2017; Liu, Cutcher, & Grant, 2017; Liu, Shin, & Burns, 2019), moving forward from the traditional use of surveys, case studies, interviews, and focus groups (Aaker, 1996; Keller, 1993; Lassar, Mittal, & Sharma, 1995).

In particular, we look at *brand importance* (Fronzetti Colladon, 2018), a recent construct inspired by the well-known concept of brand knowledge integrated with text mining and social network analysis techniques to examine whether this is associated with family firm revenues. As such, we introduce brand importance in the family business context to posit our first hypothesis. Then, we introduce the concept of family-firm identity and our second hypothesis. Specifically, we contend that the identity link between the family and the firm that characterizes some family firms is an important factor that might affect the relation between brand importance and revenues.

*2.1. Brand importance at the intersection of the family and business systems*

Family firm brands are developed at the intersection of two idiosyncratic systems, the family and the business (Astrachan et al., 2018; Craig, Dibrell, & Davis, 2008). Given their uniqueness associated with the "family nature" of the firm, family firm brands are regarded as sources of differentiation and distinctiveness, leading to obtaining a competitive advantage (Craig et al., 2008; Zellweger et al., 2010). Brands are also considered a source of heterogeneity among family firms (Blombäck & Botero, 2013; Krappe, Goutas, & von Schlippe, 2011). For instance, in their study on the branding strategies of 92 of the world's oldest family businesses, Micelotta and Raynard (2011) reveal that family firms differ in the extent to which they leverage and communicate both the family and the corporate heritage.

Studies on family business branding outcomes usually focus on the comparison of stakeholder perceptions of family versus nonfamily brands, showing that respondents positively value family ownership, control, and involvement in the firm (Orth & Green, 2009). Consumers have higher product and service expectations when buying from family firms (Carrigan & Buckley, 2008), considered superior in terms of relational qualities (Binz et al., 2013), overall reputation, and social and environmental responsibility (Sageder et al., 2015). In addition, studies show a positive link between promoting family firm brands and financial performance. This positive relationship is enhanced by an intense customer and quality orientation rooted in the strong need to protect the family name (Kashmiri & Mahajan, 2010) and the desire to create a distinctive family firm image (Zellweger et al., 2012b) by leveraging the family history, values, and identity (De Massis, Frattini, Kotlar, Petruzzelli, & Wright, 2016; Gallucci, Santulli, & Calabrò, 2015).

Past research has mainly paid attention to the differences between family and nonfamily firms, highlighting the benefits of a family nature in terms of marketing- and branding-related performance. However, an in-depth view of the contributions of external stakeholders to adding value to the brand is largely lacking. To our best knowledge, no study measures the importance that external audiences attribute to family firm brands. Yet, understanding the importance of a brand in terms of the awareness and structure of its associations as conveyed by external stakeholders is essential to advance family business research. Journalists, for example, can influence their readers and what consumers think about a brand. Brand importance is a relatively new construct comprising the three dimensions of prevalence, diversity, and connectivity as defined in Table 1 below (Fronzetti Colladon, 2018). These dimensions are related to the well-known brand knowledge and equity models (Keller, 1993; Wood, 2000). In detail, *prevalence* points to how often a brand name is mentioned in a discourse, capturing its visibility, and offering an indication of its awareness (Aaker, 1996; Keller, 1993). A high prevalence suggests that news readers will recall and recognize a brand. *Diversity* is linked to the concept of heterogeneity of brand associations and therefore related to brand image (Keller, 1993) capturing the variety and uniqueness of words mentioned in association with a brand. Heterogeneous associations are usually preferred, as they show the brand is embedded in a richer discourse (Fronzetti Colladon, 2018), contributing to brand strength (Grohs et al., 2016). The third dimension, *connectivity,* represents the extent to which a brand can bridge connections between words that are not directly connected. As Fronzetti Colladon (2018, p. 152) highlights, "connectivity could be considered as the 'brokerage' power of a brand, i.e. its ability of being in-between different groups of words, sometimes representing specific discourse topics". In prior research, some implementations of connectivity are associated with the brand popularity construct, supporting the prediction of firm financial performance (Gloor, Krauss, Nann, Fischbach, & Schoder, 2009).

**Table 1.** Definition of brand importance and its components

| Concept | Definition |
| --- | --- |
| Brand importance | The relevance a brand has in a discourse given the richness and uniqueness of its image, its visibility, and the possibility to act as a bridge connecting different discourse topics. |
| Prevalence | How frequently a brand is mentioned in a discourse (the higher the frequency, the higher the prevalence). |
| Diversity | How much a brand is associated with heterogeneous and unique words in a discourse (the richer the discourse, the higher the lexical diversity). |
| Connectivity | How frequently a brand bridges connections between words that are not directly connected (the higher the number of bridging connections, the higher the brand's connective power). |

We claim that the assessment of brand importance can offer useful insights to family firm brand managers. Indeed, when stakeholders (e.g., journalists) associate the family firm brand with heterogeneous and distinctive words, a multifaceted image is conveyed to readers (potentially customers) (Fronzetti Colladon, 2018; Grohs et al., 2016). In addition, the family firm brand's level of brokerage embeddedness (i.e., connectivity) further contributes to increasing brand importance.

Overall, these arguments suggest that a family firm whose brand is frequently mentioned in the media (prevalence), embedded in a rich and distinctive discourse (diversity), and connecting different discourse topics (connectivity), has a greater competitive advantage over other family firms in terms of the ability to attract stakeholder attention. Specifically, brand importance might attract and/or affect current and potential customer purchasing choices, with more customers buying from firms with higher brand importance, in turn leading to higher revenues. Accordingly, we posit that brand importance is associated with firm performance in terms of revenues:

**Hypothesis 1.** In family firms, brand importance is positively associated with revenues.

*2.2. Family-firm identity*

Family-firm identity, traditionally conceptualized as the degree to which the family identifies with the firm and sees the firm as an extension of the family (Chrisman et al., 2005; Zellweger et al., 2010), is considered one of the drivers of heterogeneity among family firms, with the potential to influence their goals and behaviors (Brinkerink, Rondi, Benedetti, & Arzubiaga, 2020; De Massis et al., 2018; Sundaramurthy & Kreiner, 2008; Zellweger et al., 2010). For instance, family firms with a particularly strong identity link between the family and the firm – the family is strongly involved in the business, and/or the family and the business share many of the same goals, values, beliefs, norms and interaction styles – are more likely to "acknowledge their family-owned status in their marketing and advertising" (Sundaramurthy & Kreiner, 2008, p. 419).

A common way of capturing the degree of family and firm identity overlap is by determining whether the name of the firm includes the name of the family (Deephouse & Jaskiewicz, 2013; Dyer & Whetten, 2006; Sundaramurthy & Kreiner, 2008; Zellweger et al., 2010). Family and firm name overlap is indeed one of the most powerful means to demonstrate the connection between the family and firm identity (Deephouse & Jaskiewicz, 2013; Rousseau, Kellermanns, Zellweger, & Beck, 2018), affecting both corporate reputation (Deephouse & Jaskiewicz, 2013) and transparency (Drago, Ginesti, Pongelli, & Sciascia, 2018). As a constant reminder of the connectedness and interdependence between the family and firm, the family-firm identity creates expectations of responsible firm behavior (Dyer & Whetten, 2006), since the family strongly shares its "fame or shame" (Rousseau et al., 2018, p. 11). However, with this high overlap between the family and the firm also comes intense pressure to protect and preserve the traditions and the past (De Massis et al., 2016; Erdogan, Rondi, & De Massis, 2020), emphasizing that the firm is consistent, reliable, and stable (Micelotta & Raynard, 2011).

Research suggests that when there is high identity overlap, a family firm is more likely to prioritize family aspirations over those of the business (Sundaramurthy & Kreiner, 2008), with a strong desire to preserve the values of the past and traditions (Micelotta & Raynard, 2011), attaining strategic advantages as a result of family involvement in the business (Carrigan & Buckley, 2008; Habbershon & Williams, 1999). Extending this rationale to the marketing field, several studies have shown that protecting the "family brand name" and leveraging a family-based brand identity help family firms develop social capital, positively influencing external stakeholder perceptions and even persuading consumers to make purchasing decisions based on the values, beliefs, and norms they attribute to family-owned businesses (Craig et al., 2008; Dyer, 2006). This, however, depends on the extent to which the family is able to personify the business (Miller & Le Breton-Miller, 2003), such that positive attributes associated with the family are similarly attributed to the business, and vice-versa. As a result, family-firm identity overlap may alter the relationship between the importance that external stakeholders attribute to the family firm brand and firm performance, meaning that some family firms are able to benefit more from their familiar nature than others.

Research suggests that firms lacking this overlap (i.e., separate firm and family identities) place more emphasis on business needs (Barnett, Eddleston, & Kellermanns, 2009), and family members are more likely to be seen as serving the firm rather than the firm serving the family (Sundaramurthy & Kreiner, 2008). We contend that an emphasis on the family firm's business component over the familial component counterbalances the positive effect of brand importance on firm revenues. This is because the lack of family and firm identity overlap goes against the expectations of external stakeholders, hence losing part of their appeal and struggling in developing an emotional connection with external stakeholders (Micelotta & Raynard, 2011). In other words, we expect that while brand importance is likely related to higher revenues, the lack of family and firm identity overlap might be associated with the opposite trend, downplaying the positive effect of brand importance. It follows that the firm's freedom to develop its own identity that is separate from the family's identity might be detrimental in terms of customer emotional bonds, affect, and thus orientation toward the brand. This in turn negatively affects the positive association between brand importance and firm revenues.

Conversely, we expect strong identification of the family with the firm to enhance this positive relationship. Family and firm identity overlap is traditionally associated with a plethora of positive outcomes, such as positive consumer attitudes and perceptions (Alonso-Dos-Santos & Llanos-Contreras, 2019). Family firms with strong identity overlap are usually better able to gain customer approval due to the goodwill and trustworthiness deriving from the family name, higher customer-orientation, and social responsibility (Kashmiri & Mahajan, 2010). By meeting (potential) customer expectations, family firms become more appealing in their eyes and more likely develop an emotional connection (Micelotta & Raynard, 2011). We contend this might reinforce the proposed positive relationship between brand importance and firm revenues. In other words, the positive aspects of family-firm identity overlap enhance the benefits of having an important brand in terms of revenues. Based on this reasoning, we posit:

**Hypothesis 2.** Family-firm identity overlap positively moderates the positive relationship between brand importance and revenues.

Figure 1 summarizes our hypotheses.

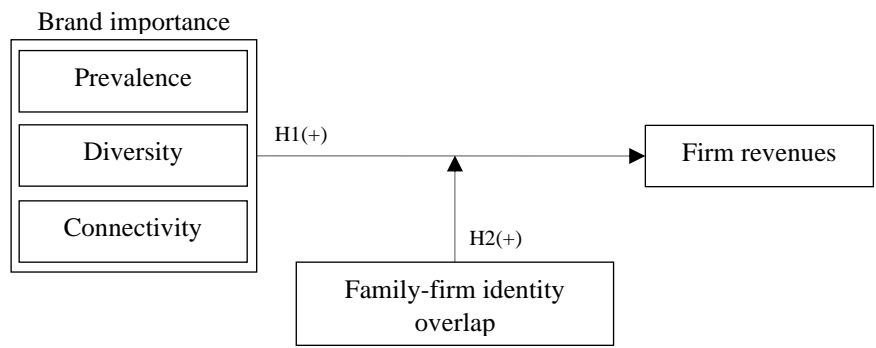

**Figure 1**. Study hypotheses.

## 3. Research methodology

### 3.1. Data collection

To test our hypotheses, we took inspiration from previous studies (e.g., Micelotta & Raynard, 2011) and derived a sample of family firms listed in the Forbes 2018 ranking of the Top 100 Italian entrepreneurial families and their businesses[1]. This type of firm is particularly relevant to our study, as entrepreneurial families (and their businesses) are renowned for their entrepreneurial orientation (Sieger, Zellweger, Nason, & Clinton, 2011; Zellweger, Kellermanns, Chrisman, & Chua, 2012a), in turn affecting their orientation toward their brand (e.g., Chang, Wang, & Arnett, 2018). From the starting list, we excluded 32 family firms either because the brand name is easily associated with famous individuals or products other than the firm, or because more than one firm has the same name (overlapping names might bias the analysis, as it was not always possible to determine automatically whether a specific word (i.e., brand name) referred to the brand). To gather data on firm characteristics, we coded information from secondary sources (e.g., their websites) and retrieved the firms' balance sheets from the Bureau van Dijk AIDA database. To assess family firm brand importance we used the semantic brand score indicator (Fronzetti Colladon, 2018) applied to the textual data of Italian online news articles published in the year 2017. Telpress International B.V. provided the news data analyzed in this study, consisting of online articles of major newspapers and news agencies in Italy. We considered all the articles that at least once mentioned the firms in our sample for a total 52,555 documents. As we excluded 5 other firms due to missing balance sheets, our final sample consists of 63 firms.

---

[1] https://forbes.it/classifica/100-famiglie-imprenditoriali-italiane-forbes/

*3.2. Study variables*

Our dependent variable is *firm revenues* in 2017, logarithmically transformed due to skewness. The main independent variable is *brand importance*, measured by means of the SBS indicator. The SBS is a novel composite indicator applicable to any textual data, calculated by combining text mining and social network analysis methods and tools (Fronzetti Colladon, 2018). The three dimensions are brand *prevalence*, *diversity*, and *connectivity*. Prevalence measures the frequency of occurrence of the brand name, assuming that brands that are named more frequently are more important, as they have generated higher awareness (Keller, 1993) in terms of both the writer's and the reader's perspective. However, prevalence alone is not sufficient to measure brand importance. Indeed, a brand name might be mentioned very frequently, but always in association with the same low-informative words. Therefore, the diversity dimension takes into account the heterogeneity and uniqueness of the textual brand association. To calculate diversity, we constructed a social network graph based on the co-occurrence of words in the text. The g-graph consists of $n$ nodes and $m$ edges, where each word appearing in the text is a node. The arcs interconnecting the nodes are weighted according to the frequency of the co-occurrence of every node pair. We took 5 as a threshold for the maximum co-occurrence distance, as suggested in past research, proving the SBS results are robust to variations of this parameter[2] (Fronzetti Colladon, 2018). In addition, we filtered out rare co-occurrences (i.e., links with low weights). Based on this graph, we calculated the distinctiveness centrality metric to measure diversity (Fronzetti Colladon & Naldi, 2020). Distinctiveness is higher when a brand (node) has more links (i.e., higher number of associations), and when these associations are less common. Finally, the third connectivity dimension reflects the brand's ability to act as a bridge, connecting other words and ultimately discourse topics. Connectivity measures the "brokerage power" of the analyzed brand in the co-occurrence network, operationalized through the metric of weighted betweenness centrality (Wasserman & Faust, 1994). The SBS indicator results from the sum of the standardized measures of prevalence, diversity, and connectivity.

Prior to calculating the SBS, we preprocessed (Perkins, 2014) the news data (i.e., documents) to remove: (1) words that add little meaning to the text (stop-words, e.g., "and", "or"); (2) word affixes (known as stemming) (Porter, 2006); and (3) punctuation and special characters. For all the natural language processing, brand associations, and SBS computation tasks, we used the SBS BI webapp[3] (Fronzetti Colladon & Grippa, 2020). The computing resources were provided by the Italian National Agency for New Technologies, Energy and Sustainable Economic Development (ENEA), as we used a version of the app hosted on the ENEA/CRESCO infrastructure (Iannone et al., 2019).

To assess the moderating effect of family-firm identity, we created the dummy variable *family and firm name overlap*, which reflects the level of identification of the family with the firm (e.g., Pérez-González, 2006; Villalonga & Amit, 2006). Following Deephouse and Jaskiewicz (2013) and Rousseau et al. (2018), the dummy variable is equal to 1 when the family name is included in the firm name, 0 otherwise.

Finally, we considered several control variables taken from the literature (e.g., De Massis, Eddleston, & Rovelli, 2021; Rondi & Rovelli, 2021). *Family CEO* is a dummy variable equal to 1 if the CEO is a member of the owning family. *Firm generation* indicates the generation that is managing the firm. *Generations involved* is a dummy variable equal to 1 if more than one family generation is involved in the management of the firm, 0 otherwise. We also considered *firm age* and *firm size,* i.e., number of employees. We then added a series of dummy variables: (1) *geographic area dummies*, which indicate whether the family firm is located in the north-east, north-west, center, or south of Italy; (2) *industry dummies*, which indicate whether the firm operates in manufacturing, services, or constructions; (3) *legal status dummies*, which indicate the legal status of the firm (i.e., joint-stock company, limited liability company, others). Finally, we controlled for the *sentiment* of news data to understand whether firm revenues are associated not only with brand importance, but also with how journalists talk about the firm. We calculated brand sentiment through the SBS BI app (Fronzetti Colladon & Grippa, 2020), with scores varying in the range [-1,1] and negative values suggesting negative feelings about the brand.

---

[2] We also repeated our analysis with a different threshold (i.e., 7), and in line with previous studies (e.g., Fronzetti Colladon, 2018), the results did not significantly differ.

[3] https://bi.semanticbrandscore.com

## 4. Results

Table 2 presents the descriptive statistics and correlations.

**Table 2.** Descriptive statistics and correlations (*p*-values in parentheses).

|     |                              | Mean    | S.D.    | (1)     | (2)     | (3)     | (4)     | (5)     | (6)     | (7)     | (8)     | (9)   |
|-----|------------------------------|---------|---------|---------|---------|---------|---------|---------|---------|---------|---------|-------|
| (1) | Firm revenues                | 1189130 | 3784905 | 1.000   |         |         |         |         |         |         |         |       |
| (2) | Brand importance             | 0.693   | 4.689   | 0.489   | 1.000   |         |         |         |         |         |         |       |
|     |                              |         |         | (0.000) |         |         |         |         |         |         |         |       |
| (3) | Family and firm name overlap | 0.603   | 0.493   | -0.056  | -0.132  | 1.000   |         |         |         |         |         |       |
|     |                              |         |         | (0.662) | (0.302) |         |         |         |         |         |         |       |
| (4) | Family CEO                   | 0.603   | 0.493   | -0.244  | -0.344  | 0.115   | 1.000   |         |         |         |         |       |
|     |                              |         |         | (0.054) | (0.006) | (0.607) |         |         |         |         |         |       |
| (5) | Firm generation              | 4.286   | 5.754   | -0.221  | -0.118  | 0.209   | 0.029   | 1.000   |         |         |         |       |
|     |                              |         |         | (0.082) | (0.357) | (0.100) | (0.822) |         |         |         |         |       |
| (6) | Generations involved         | 0.603   | 0.493   | 0.216   | -0.099  | 0.005   | 0.219   | 0.001   | 1.000   |         |         |       |
|     |                              |         |         | (0.090) | (0.441) | (0.967) | (0.303) | (0.995) |         |         |         |       |
| (7) | Firm size                    | 2159.175| 5465.239| 0.510   | 0.348   | -0.222  | -0.200  | -0.092  | -0.236  | 1.000   |         |       |
|     |                              |         |         | (0.000) | (0.005) | (0.081) | (0.117) | (0.472) | (0.063) |         |         |       |
| (8) | Firm age                     | 103.667 | 111.005 | -0.084  | 0.024   | 0.215   | -0.059  | 0.792   | 0.070   | -0.028  | 1.000   |       |
|     |                              |         |         | (0.515) | (0.851) | (0.091) | (0.647) | (0.000) | (0.587) | (0.826) |         |       |
| (9) | Sentiment                    | 0.107   | 0.042   | 0.211   | 0.249   | -0.033  | -0.299  | -0.120  | 0.248   | -0.106  | -0.024  | 1.000 |
|     |                              |         |         | (0.097) | (0.049) | (0.797) | (0.017) | (0.348) | (0.050) | (0.407) | (0.851) |       |

Notes: We computed the correlations considering non-transformed and standardized variables. For correlations between dummy variables, we computed the tetrachoric correlation, and between the dummy and continuous variables, we computed the point biserial correlation.

As for our main study variables, *brand importance* is positively and significantly correlated with *firm revenues* (rho = 0.489, *p*-value = 0.000), providing a first hint of the possible significance of the relationship posited in H1. *Brand importance* is also positively correlated with *firm size* (rho = 0.348, *p*-value = 0.005) and *sentiment* (rho = 0.249, *p*-value = 0.049), but negatively correlated with *family CEO* (rho = -0.344, *p*-value = 0.006). On the other hand, *family and firm name overlap* is not significantly correlated with either *firm revenues* or *brand importance*. Instead, it is positively correlated with older (*firm age*, rho = 0.215, *p*-value = 0.091) and smaller firms (*firm size*, rho = -0.222, *p*-value = 0.081). Finally, *firm revenues* is negatively correlated with *family CEO* (rho = -0.244, *p*-value = 0.054) and *firm generation* (rho = -0.221, *p*-value = 0.082), and positively correlated with *generations involved* (rho = 0.216, *p*-value = 0.090), *firm size* (rho = 0.510, *p*-value = 0.000), and *sentiment* (rho = 0.211, *p*-value = 0.097).

Table 3 shows the results of the OLS models we used to test our hypotheses. To exclude multicollinearity, we performed variance inflation factor (VIF) tests. The maximum VIF is 2.18, and the average VIF is 1.86, which are both lower than the thresholds generally associated with multicollinearity problems (Belsley, Kuh, & Welsch, 1980). Before running the models, we standardized all continuous variables to ease the comparison of the resulting coefficients.

**Table 3**. OLS models testing the relation between *brand importance* and *firm revenues*, and the moderating effect of *family and firm name overlap* (dependent variable *firm revenues*).

|  | Model 1 | Model 2 | Model 3 | Model 4 |
|---|---|---|---|---|
| Brand importance (SBS) | - | 0.625*** | 0.628*** | 0.351** |
|  |  | (0.130) | (0.123) | (0.154) |
| Family and firm name overlap | - | - | 0.141 | 0.086 |
|  |  |  | (0.174) | (0.172) |
| Brand importance (SBS) * Family and firm name overlap | - | - | - | 0.502** |
|  |  |  |  | (0.211) |
| Family CEO | -0.324 | -0.150 | -0.155 | -0.071 |
|  | (0.240) | (0.206) | (0.203) | (0.198) |
| Firm generation | -0.244** | -0.114 | -0.113 | -0.160** |
|  | (0.101) | (0.079) | (0.077) | (0.070) |
| Generations involved | 0.696*** | 0.706*** | 0.713*** | 0.597*** |
|  | (0.227) | (0.185) | (0.179) | (0.197) |
| Firm size | 0.502*** | 0.341** | 0.355** | 0.408*** |
|  | (0.123) | (0.134) | (0.138) | (0.135) |
| Firm age | 0.108 | -0.007 | -0.024 | 0.036 |
|  | (0.103) | (0.075) | (0.073) | (0.072) |
| Sentiment | 1.683 | 0.171 | 0.260 | 0.248 |
|  | (2.396) | (2.018) | (2.098) | (2.070) |
| Geographical area dummies | YES | YES | YES | YES |
| Industry dummies | YES | YES | YES | YES |
| Legal status dummies | YES | YES | YES | YES |
| Constant | 1.902** | 2.867*** | 2.707*** | 2.654*** |
|  | (0.827) | (0.735) | (0.844) | (0.802) |
| Observations | 63 | 63 | 63 | 63 |
| Log-likelihood | -61.26 | -49.88 | -49.43 | -46.84 |
| R-squared | 0.584 | 0.710 | 0.714 | 0.737 |

\* $p < 0.1$, ** $p < 0.05$, ***$p < 0.01$. Robust standard errors in parentheses.

Model 1 is the baseline model including only the control variables. In Model 2, we added the independent variable *brand importance*, while Model 3 also considers the effect of *family and firm* *name overlap*. In Model 4, we tested the moderating effect of *family and firm name overlap* on the relationship between *brand importance* and *firm revenues*. To interpret the results of the moderation, we used the Delta method (Hoetker, 2007) and computed the average marginal effects (AMEs).

Model 1 indicates that *firm revenues* increase with *firm size* (coef = 0.502, *p*-value = 0.000) and with the involvement of more than one generation in the firm's management (*generations involved*, coef = 0.696, *p*-value = 0.004), while they decrease with *firm generation* (coef = -0.244, *p*-value = 0.020). Interestingly, *sentiment* is not significantly associated with *firm revenues*, suggesting that the way in which external stakeholders talk about the firm might not play a decisive role in driving performance. In Model 2, we introduce our main independent variable. Confirming H1, *brand importance* is significantly associated with *firm revenues* (coef = 0.625, *p*-value = 0.000). This means that the greater the brand importance generated by external stakeholders, the higher the revenues. More specifically, one standard deviation increase in *brand importance* is associated with a 0.625 percentage point increase in *firm revenues*. This highly significant result still holds (coef = 0.628, *p*-value = 0.000) when including *firm and firm name overlap* in Model 3, which instead is not significantly related to revenues.

Model 4 supports H2 in that *firm and firm name overlap* significantly and positively moderates (*p*-value = 0.021) the association between *brand importance* and *firm revenues*. Specifically, the AMEs reported in Table 4 reveal that in family firms, *brand importance* is more positively and significantly related to *firm revenues* when there is *family and firm name overlap*. Indeed, the AME of *brand importance* is of greater magnitude (and significance) when the family identifies with the firm (AME = 0.853, *p*-value = 0.000). This result is shown in Figure 2 and confirms H2 positing that in family firms, family and firm name overlap positively moderates the relationship between brand importance and firm revenues. In other words, the positive relation between brand importance and revenues is stronger when the family identifies with its firm.

**Table 4**. Average marginal effects of *brand importance* at two levels of *family and firm name overlap*.

|                                   | **AME of brand importance (SBS)** | *p*-value | t-test (*p*-value) |
|-----------------------------------|:---:|:---:|:---:|
| Family and firm name overlap = 0  | 0.351 | 0.027 |  |
|                                   | (0.154) |  | 0.021 |
| Family and firm name overlap = 1  | 0.853 | 0.000 |  |
|                                   | (0.142) |  |  |

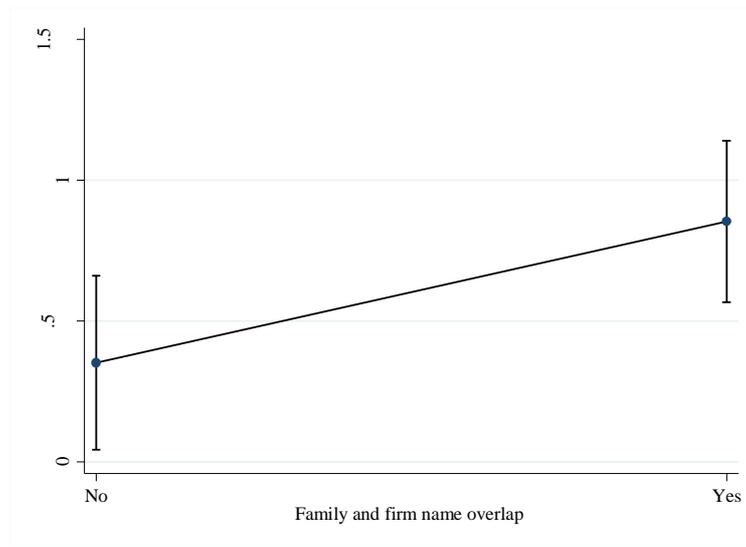

**Figure 2**. Average marginal effect of brand importance at the two levels of family and firm name overlap (95% confidence interval).

Looking more closely at the relationship between *family and firm name overlap* and *firm revenues*, which past research suggests is positive (Craig et al., 2008; Dyer, 2006), we find that it is indeed positive, but only above a certain threshold of *brand importance* (Figure 3). This indicates that a minimum level of *brand importance* (generated by external stakeholders) is needed to enable the positive power of family-firm identity.

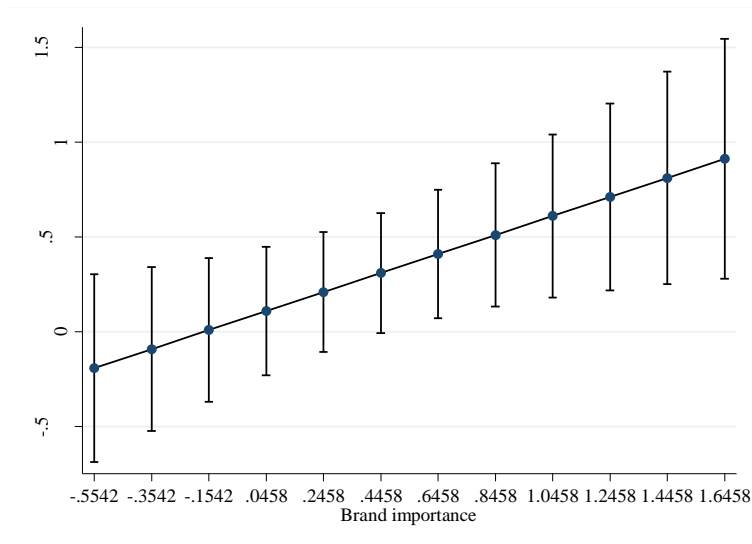

**Figure 3**. Average marginal effect of family and firm name overlap at increasing levels of brand importance (95% confidence interval).

## 5. Discussion and conclusions

In this research, we study family firm brands by adopting an external perspective and exploring the relationship between brand importance and firm revenues. In line with previous studies on family firm marketing and branding (Micelotta & Raynard, 2011), we analyzed a sample of 63 of the top 100 Italian entrepreneurial families and their businesses. We assessed these firms' brand importance through text mining and social network analysis techniques, i.e., the SBS indicator (Fronzetti Colladon, 2018), considering 52,555 Italian online news articles about their brands. Results show that brand importance is positively and significantly associated with firm revenues. Our study suggests that a family firm whose brand importance is fostered by external audiences – meaning that the brand is frequently mentioned in the media (prevalence), is embedded in a rich and distinctive discourse (diversity), and connects different discourse topics (connectivity) – obtains a competitive advantage, which likely translates into higher revenues. Nevertheless, our findings also reveal that this positive relation is contingent on a particular characteristic and source of heterogeneity among family firms, namely the identity link between the family and the firm. Specifically, the positive association between brand importance and firm revenues is of greater magnitude when the family identifies with the firm. This could be explained by consumers' emotional bonds, affect, and orientation toward the brand resulting from the firm highlighting the familial component. Promoting the family background adds to the positive effect of high brand importance. Moreover, identity overlap plays a significant role only when brand importance is sufficiently high.

Our study advances family business research on branding by introducing the new concept of brand importance, which integrates traditional approaches with a more detailed consideration of a brand's visibility, its embeddedness in a rich and distinctive discourse, and connections with different discourse topics. Moreover, adopting an external perspective through collecting different media sources we shed new light on the external perception of family firm brands in a way that acknowledges not only their heterogeneity but also the heterogeneity of external audiences, taking into account the media as a different group of external stakeholders (Beck, 2016). Our study offers a deeper understanding of how brand importance might drive external audiences when confronting family firm brands. Assessing brand importance provides family firms with the opportunity to gain a richer perspective in evaluating their brand to understand its relationship with performance.

Furthermore, contributing to family business research (Rovelli, Ferraso, De Massis, & Kraus, 2021), we show that family-firm identity affects the relationship between family firm brand importance and performance. We thus extend prior studies acknowledging that the overlap between family and firm identity is an important determinant of family firm behavior (e.g., Brinkerink, Rondi, Benedetti, & Arzubiaga, 2020; De Massis et al., 2018; Sundaramurthy & Kreiner, 2008; Zellweger et al., 2010), and may explain family firm heterogeneity (Chua, Chrisman, Steier, & Rau 2012). Compared to prior research that assumes low levels of self-complexity, focusing on the effects of a single identity rather than the additive or multiplicative effects of several identities (Creary, Caza, & Roberts, 2015; Ramarajan, 2014; Ramarajan, Rothbard, & Wilk, 2017), our study recognizes organizational identity in family firms as a multifaceted and dynamic construct. With our approach, we unveil how family firms vary in their level of subjective self-complexity based on the degree to which the family and firm identity overlaps (Linville, 1985; Roccas & Brewer, 2002).

Moreover, by applying SBS to study family firm brands, we move forward from the traditional use of surveys, case studies, interviews, or focus groups (Aaker, 1996; Keller, 1993; Lassar et al., 1995), providing a new (big data) approach based on the discourse analysis of a considerable number of online news articles. Our method allows repeatable and automated measurements to continuously monitor brand importance. We believe that our research exploits the opportunities offered by the availability of rich online text data and provides additional evidence that can inspire researchers in the family business field. Likewise, practitioners could be encouraged to adopt big data methods to support their decision-making processes. Last, we extend research on brand importance and the application of SBS to the family business context.

It is worth noting that our findings partially contrast with studies attributing high importance to the positivity of messages for the prediction of consumer behavior (e.g., Kim & Ko, 2012), as in our setting sentiment was mostly non-influential. This is aligned with past research focused, for example, on the analysis of museum brands and the non-significant impact of sentiment on museum visitors (Fronzetti Colladon, Grippa, & Innarella, 2020). Therefore, while it is important to monitor the sentiment of external stakeholders towards a brand, our results suggest that family business scholars should pay more attention to the importance of the brand rather than the sentiment it generates.

Our study has some limitations that open up opportunities for future research. First, online news are not always easily and freely available, especially in the case of massive downloads. Telpress International supported our data collection by providing its dataset of articles for 2017, hence limiting our study to be cross-sectional. Accordingly, we do not claim causality, but provide evidence of the significant relationships among our variables of interest. While we are confident of the quality and reliability of our findings, we invite scholars to replicate our methodology over a longer period. Indeed, a longitudinal study would be useful to gain a deeper understanding of the investigated relationships. Second, our empirical investigation relies on data from Italy, which might also limit the generalizability of our results, as the relationship between brand importance and revenues, and the moderating effect of family-firm identity might change depending on the cultural context in which the family firm operates. Future studies might consider different countries.

Third, in our analysis, we were not able to disentangle alternative types of external media sources. While we analyzed textual data from online articles from different sources (i.e., newspapers and news agencies), it was not possible to determine whether the relationship between family firms' brand importance and revenues, and the moderating effect of family-firm identity depend on the type of external media writing and publishing the articles. We hence encourage scholars to explore the role of different media sources in the relationship between brand importance and firm revenues, and obtain deeper insights on the heterogeneity of the perceptions of external stakeholders toward family firm brands. Likewise, we encourage scholars to take a situational approach to study how external audiences perceptions of the brand shape family firm performance. For instance, scholars might examine whether the role played by external audiences in determining the importance of family firm brands and the relationship with firm performance vary in the imminence of restructuring initiatives with high impact on the media, such as mergers, acquisitions, spin-offs, sell-offs or buy-outs (King et al., 2021).

Despite these limitations, our study provides some practical suggestions for family firm managers. First, our results show the relevance of using brand importance – a measure that considers prevalence, diversity, and connectivity – and demonstrate that the importance that external audiences attach to family brands is positively related to revenues. This suggests that family firms should invest in their relationships with external audiences (e.g., media, consumers, and other stakeholders). Given the positive relationship between brand importance and higher revenues in the case of family identification with the firm, we suggest that family firms carefully choose whether they strategically stress the link between the family and the firm in their branding activities. In this vein, family firms might benefit from using SBS and the BI tool to analyze the importance and image of their brand to make more informed marketing and branding decisions. Second, our study shows that the sentiment that news generate about the family firm brand does not relate to revenues. We thus encourage family firms to be less concerned about the positive or negative feelings conveyed by news and instead consider enhancing their brand's importance.

Chang, Y., Wang, X., & Arnett, D. B. (2018). Enhancing firm performance: The role of brand orientation in business-to-business marketing. *Industrial Marketing Management, 72*, 17–25.

Chrisman, J. J., Chua, J. H., & Sharma, P. (2005). Trends and directions in the development of a strategic management theory of the family firm. *Entrepreneurship Theory and Practice, 29*(5), 555–575.

Chua, J. H., Chrisman, J. J., Steier, l. P., & Rau, S. B. (2012). Sources of heterogeneity in family firms: An introduction. *Entrepreneurship Theory and Practice, 36*(6), 1103–1113.

Craig, J. B., Dibrell, C., & Davis, P. S. (2008). Leveraging family-based brand identity to enhance firm competitiveness and performance in family businesses. *Journal of Small Business Management, 46*(3), 351–371.

Creary, S. J., Caza, B. B., & Roberts, L. M. (2015). Out of the box? How managing a subordinate's multiple identities affects the quality of a manager-subordinate relationship. *Academy of Management Review, 40*(4), 538–562.

De Massis, A., Frattini, F., Kotlar, J., Petruzzelli, A. M., & Wright, M. (2016). Innovation through tradition: Lessons from innovative family businesses and directions for future research. *Academy of Management Perspectives, 30*(1), 93–116.

De Massis, A., Eddleston, K. A., & Rovelli, P. (2021). Entrepreneurial by Design: How Organizational Design Affects Family and Nonfamily Firms' Opportunity Exploitation. *Journal of Management Studies*, *58*(1), 27–62.

De Massis, A., Kotlar, J., Mazzola, P., Minola, T., & Sciascia, S. (2018). Conflicting selves: Family owners' multiple goals and self-control agency problems in private firms. *Entrepreneurship Theory and Practice, 42*(3), 362–389.

Deephouse, D. L. (2000). Media reputation as a strategic resource: An integration of mass communication and resource-based theories. *Journal of Management, 26*(6), 1091–1112.

Deephouse, D. L., & Jaskiewicz, P. (2013). Do family firms have better reputations than non-family firms? An integration of socioemotional wealth and social identity theories. *Journal of Management Studies, 50*(3), 337–360.

Drago, C., Ginesti, G., Pongelli, C., & Sciascia, S. (2018). Reporting strategies: What makes family firms beat around the bush? Family-related antecedents of annual report readability. *Journal of Family Business Strategy, 9*(2), 142–150.

Dyer, W. (2006). Examining the "family effect" on firm performance. *Family Business Review, 19*(4), 253–273.

Dyer, W. G., & Whetten, D. A. (2006). Family firms and social responsibility: Preliminary evidence from the S&P 500. *Entrepreneurship Theory and Practice, 30*(6), 785–802.

Erdogan, I., Rondi, E., & De Massis, A. (2020). Managing the tradition and innovation paradox in family firms: A family imprinting perspective. *Entrepreneurship Theory and Practice*, *44*(1), 20–54.

Fronzetti Colladon, A. (2018). The semantic brand score. *Journal of Business Research, 88*, 150–160.

Fronzetti Colladon, A. (2020). Forecasting election results by studying brand importance in online news. *International Journal of Forecasting, 36*(2), 414–427.

Fronzetti Colladon, A., & Grippa, F. (2020). Brand intelligence analytics. In A. Przegalinska, F. Grippa, & P. A. Gloor (Eds.), *Digital Transformation of Collaboration* (pp. 125–141). Cham, Switzerland: Springer Nature Switzerland.
26